\documentclass[twocolumn,superscriptaddress,amsmath,amssymb]{revtex4-1}
\usepackage[english]{babel}
\usepackage[combine]{ucs}
\usepackage[utf8x]{inputenc}
\usepackage[dvips]{graphicx}
\usepackage{amssymb}

\begin{document}

\author{E. H. {\AA}hlgren}
\affiliation{Department of Physics, University of Helsinki, P.O. Box
  43, 00014 Helsinki, Finland}
\author{J. Kotakoski}
\email[Corresponding author. E-mail: ]{jani.kotakoski@iki.fi}
\affiliation{Department of Physics, University of Helsinki, P.O. Box
  43, 00014 Helsinki, Finland}
\affiliation{Department of Physics, University of Vienna,
  Boltzmanngasse 5, 1190 Wien, Austria}
\author{O. Lehtinen}
\affiliation{Department of Physics, University of Helsinki, P.O. Box
  43, 00014 Helsinki, Finland}
\author{A.~V. Krasheninnikov}
\affiliation{Department of Physics, University of Helsinki, P.O. Box
  43, 00014 Helsinki, Finland}
\affiliation{Department of Applied Physics, Aalto University, P.O. Box
  1100, 00076 Aalto, Finland}

\title{Ion irradiation tolerance of graphene as studied by atomistic
  simulations}

\begin{abstract}
As impermeable to gas molecules and at the same time transparent to high-energy
ions, graphene has been suggested as a window material for separating a
high-vacuum ion beam system from targets kept at ambient conditions. However,
accumulation of irradiation-induced damage in the graphene membrane may give
rise to its mechanical failure. Using atomistic simulations, we demonstrate
that irradiated graphene even with a high vacancy concentration does not show
signs of such instability, indicating a considerable robustness of graphene
windows. We further show that upper and lower estimates for the irradiation
damage in graphene can be set using a simple model.
\end{abstract}

\maketitle

Since the isolation of graphene in 2004~\cite{Novoselov2004}, a
multitude of applications have been proposed for this one-atom-thick
carbon membrane~\cite{Geim2009a}.  Although graphene is probably best
known for its unique electronic properties~\cite{CastroNeto2009}, it
is also the strongest material ever measured~\cite{Lee2008}. Moreover,
on the one hand, even a single layer of graphene can withstand
pressure imposed by a macroscopic amount of gas~\cite{Stolyarova2009},
and does not allow even the smallest atmospheric molecules to
permeate~\cite{Bunch2008}. On the other hand, porous graphene has been
considered to be the ultimate membrane for gas
separation~\cite{Jiang2009c} and an ideal material for
supercapasitors~\cite{Zhu2011a}. At the same time, graphene is virtually
transparent to high energy ions, which pass through the material without
creating substantial damage due to negligible interaction cross
section~\cite{Lehtinen2010b}. Due to the unique combination of strength and
impermeability to gases, graphene can be used as a window material in external
ion beam experiments with samples which cannot be put into vacuum required for
the operation of the ion-beam system~\cite{Lehtinen2010b}, replacing
silicon-nitride membranes~\cite{Dran2004} currently used for this purpose.
Very recently, a similar technology has been demonstrated with graphene oxide
windows for {\it in situ} environmental cell photoelectron
spectroscopy~\cite{Kolmakov2011}.

Nevertheless, irradiation damage can accumulate with increasing ion
dose, so that the operation of graphene as an ion-transparent but
gas-separating membrane depends crucially on its ability to withstand
continuous irradiation during the experiment. Although the ion
irradiation response of graphene has been recently studied both
experimentally~\cite{Tapaszto2008,Chen2009,Compagnini2009,Giannazzo2009,Zhou2010a,Mathew2011,Ney2011,Mathew2011a,Compagnini2011,Chen2011,Diaz-Pinto2011,Buchowicz2011,Cancado2011,Ugeda2011,Ugeda2012}
and theoretically~\cite{Lehtinen2010b,Ahlgren2011,Lehtinen2011b}, the
atomic-scale details of damage accumulation during continuous exposure
to the ion beam remain unknown. In the experiments, ion doses were
either very low creating only spatially well-separated defects or so
high that graphene was completely destroyed, and it was not possible
to get any insight into damage accumulation process. Similarly, the
theoretical work has hitherto concentrated on the effects of
individual ion impacts on pristine graphene. Continuous high-dose
irradiation has been taken into account only in a stochastic manner
disregarding the actual dynamics of consecutive ion impacts on a
defective graphene structure~\cite{Lehtinen2011b}.

In this Letter, we utilize atomistic simulations to study the effects
of ion irradiation on graphene with defects to understand the details
of damage accumulation.  We demonstrate that irradiated graphene with
vacancy concentration of at least 35\% does not show any signs of
structural failure, pointing to considerable stability of
graphene windows in ion beam experiments. Our results can be directly
utilized both in estimating the wear of graphene windows used to
separate ion beam systems from volatile targets as well as in
designing optimum parameters for carving nanopores into graphene
membranes using a focused ion beam.

\begin{figure}[ht!]
    \includegraphics[width=.95\linewidth]{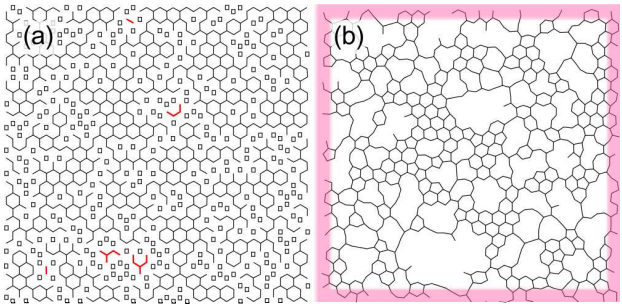}
    \caption{Overview of the simulations setup for a very high vacancy
		concentration. Bonds between the atoms are drawn with solid lines. (a) An
		example structure as created with the vacancies marked with open squares.
		Thicker (red) lines show isolated islands which were also removed from the
		target structure. (b) The structure after annealing and structural
		optimization. The wide semi-transparent lines at the boundaries indicate
		the heat dissipation area.} \label{px::overview}
  \end{figure}

We created our graphene target structures by randomly removing atoms
from the pristine lattice, then running a 3~ps annealing simulation at
800~K to allow saturation of open bonds where possible, and finally
relaxing the structure to its local energy minimum by quenching to 0~K. At
this point we also removed all isolated fragments leaving only the largest
continuous atomic network to be used as a target for our ion irradiation
simulations, as illustrated in Fig.~\ref{px::overview}. The initial structure
consisted of 1250 carbon atoms. We performed in total 200,000
irradiation simulations for individually created structures with
impact points randomly selected in the middle of the simulation
cell. For each structure, ion species (He, Ar, Xe), ion energy ($K\in
[30~\mathrm{eV},1~\mathrm{MeV}]$) and initial vacancy concentration
($c_V$) combination, we performed on average more than 1050
independent simulations in order to collect representative statistics.

Irradiation simulations were carried out using the molecular dynamics
(MD) method as implemented in the {\scshape parcas} simulation
code~\cite{Nordlund1998a}. The carbon-carbon interaction was described
by the reactive bond-order potential developed by Brenner {\it et
  al.}~\cite{Brenner1990b,Brenner1992}, disregarding the bond
conjugation term which is not important for ion irradiation
effects~\cite{Krasheninnikov2001}. The interactions between noble gas
ions~\footnote{We refer to the projectile as {\it ion} throughout the
  text although the charge of the impinging particle is not explicitly
  taken into account in our simulation setup.} and carbon atoms were
modeled by the universal repulsive potential by Ziegler, Biersack and
Littmark~\cite{Ziegler1985}. A similar repulsive potential was fitted
to the carbon-carbon interaction at short distances in order to
properly describe high energy collisions between carbon atoms. A few
atomic rows at the edges of the periodic directions were coupled to
the Berendsen thermostat~\cite{Berendsen1984} kept at 0~K in order to
model the dissipation of heat from the irradiated area, as indicated
in Fig.~\ref{px::overview}b. We have previously used a similar
simulation setup for modeling ion irradiation of carbon nanotubes
~\cite{Tolvanen2007a,Krasheninnikov2002b}, pristine
graphene~\cite{Lehtinen2010b,Ahlgren2011} and hexagonal boron nitride
mono-layers~\cite{Lehtinen2011}. After each ion impact we again quenched the
system into a local energy minimum for analysis.

\begin{figure*}[ht!]
  \includegraphics[width=.6\linewidth]{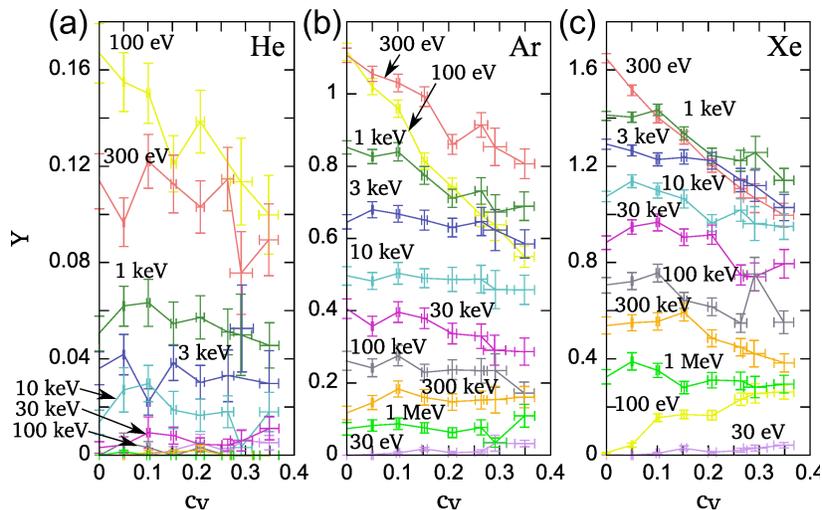}
    \caption{Sputtering yield ($Y$) as a function of vacancy
			concentration ($c_V$) for (a) He, (b) Ar and (c) Xe ions. Each data
			point is an average $Y$ for all simulations for the same ion species,
			ion energy and $c_V$. The
      error bars corresponding to $Y$ show the standard deviation and
      the error bars corresponding to $c_V$ are caused by the removal
      of unconnected fragments after randomly removing atoms from
      pristine graphene.}
    \label{px::ycvac}
  \end{figure*}

In Fig.~\ref{px::ycvac} we present sputtering yield ($Y$) for
different ions and irradiation energies as a function of $c_V$.  Note
that in the limit $c_V\rightarrow 0$, the results approach those for
pristine graphene~\cite{Lehtinen2010b}, as expected. Despite the fact that the
presented results are averaged over a large
number of simulations, significant statistical variations remain present
in the data. However, the qualitative behavior of $Y$ as a function of
vacancy concentration is nevertheless apparent. For instance, it is
clear that $Y$ tends to decrease with increasing $c_V$ due to increased
probability for the ion to pass through an existing vacancy with an
energy-dependent slope until the highest considered vacancy
concentrations ($c_V \leq 35$\%). Perhaps the most surprising result is
the continuity of all of the curves (except for some statistical
fluctuations), since one could assume that at some point the
defective membrane would become structurally unstable so that it would
break showing an abrupt increase in the sputtering yield
(corresponding to a lost membrane).  Our results show that such an
instability point is not reached within any reasonable vacancy
concentration. Indeed, a careful analysis of the distribution of the
number of sputtered atoms as a function of $c_V$ shows no abrupt
changes in the relative probabilities for sputtering smaller or larger
numbers of atoms at once. The structural stability of
the membranes was further checked by 1~ns anneal simulations at 1500~K for 150
structures with different $c_V$ after the analysis. No apparent
instabilities, e.g., disintegration or crumpling, were observed. We also point
out that although mechanical properties of perforated graphene are inferior as
compared to the pristine material, the fracture stress remains as high as
$\sim 50$~GPa corresponding to nearly 10\% strain for graphene with up to 2~nm
holes~\cite{Khare2007}.

\begin{figure}[ht!]
  \includegraphics[width=.95\linewidth]{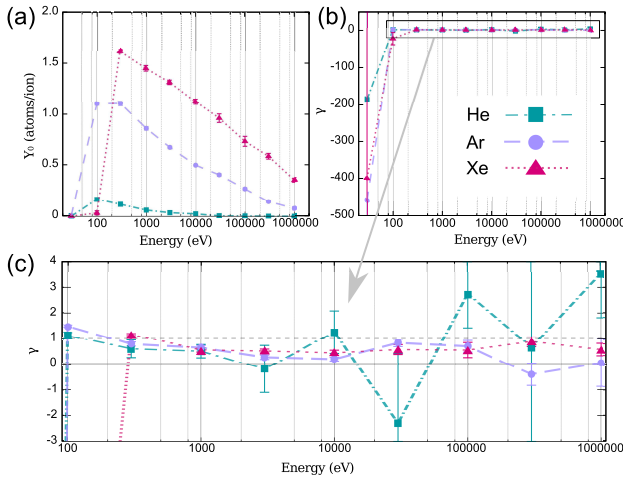}
  \caption{(a) Fitted sputtering yield $Y_0$ at zero vacancy
    concentration ($c_V=0$) and (b) fitted dimensionless constant
    $\gamma$ indicating deviation from the simple geometric
    model. Panel (c) presents a magnification of the area marked with
    a rectangle in panel (b).}
  \label{px::slopes}
\end{figure}

An additional surprise is the clearly evident ion energy-dependent
slope of the data presented in Fig.~\ref{px::ycvac}. Based purely on
geometrical arguments, $Y$ should decrease with $c_V$ due to a drop in
target density and thus smaller collision cross section as
\begin{equation}
  Y (c_V ) = (1 - c_V )Y_0,
  \label{eq::geom}
\end{equation}
where $Y_0 = Y(c_V=0)$ corresponds to pristine graphene. Since this
equation does not depend on the energy of the impinging ion, it is
clear that it cannot completely describe the data shown in
Fig.~\ref{px::ycvac}. Therefore, to quantify the differences between
the data and the geometric model, we introduced a new dimensionless
variable $\gamma$ in Eq.~\ref{eq::geom} to obtain
\begin{equation}
  Y (c_V ) = (1 - \gamma c_V )Y_0.
  \label{eq::geom2}
\end{equation}
Fitting the data presented in Fig.~\ref{px::ycvac} to
Eq.~\ref{eq::geom2} allows us to systematically analyze the deviation
from the geometric model as a function of ion energy. The results of
the fits are presented in Fig.~\ref{px::slopes}. Naturally, the
fitted sputtering yield at zero vacancy concentration ($Y_0$) for each
energy/ion combination (Fig.~\ref{px::slopes}a) is nearly identical to
the simulation data for pristine graphene~\cite{Lehtinen2010b}. The
modified equation reduces to the simple geometric model
(Eq.~\ref{eq::geom}) for $\gamma=1$, whereas negative values indicate
increasing $Y$ for increasing $c_V$ (a positive slope), and $\gamma=0$
would suggest sputtering yield which does not depend on $c_V$.

As evident from Fig.~\ref{px::slopes}(b), $\gamma$ is negative at low
ion energies. In the case of He and Ar, this occurs only up to
$K=30$~eV, whereas for the heaviest Xe also the 100~eV data has a
positive slope. For energies immediately above these, $\gamma \approx
1$ indicating a perfect agreement with the simple geometric
model. Then, $\gamma$ decreases until an apparent saturation towards a
constant value (within our statistical accuracy) at an ion-dependent
energy. At the saturation the average values of $\gamma$ are $1.08
\pm 0.62$, $0.32 \pm 0.10$ and $0.53 \pm 0.05$ for He, Ar and Xe, respectively
(notice that the statistical uncertainty decreases for increasing ion mass).

The varying values of $\gamma$ can be understood taking into account
how the collision process changes with increasing ion momentum (and
thus energy)~\cite{Lehtinen2010b}. At low energies, the collision
cross section is large, and the collision process is relatively
slow. The pre-existing vacancies in the structure on average lower the
binding energy of the target atoms which makes an important
contribution to the sputtering when the transferred energies are very
close to the displacement threshold (22~eV in pristine
graphene~\cite{Kotakoski2010,Meyer2012} and somewhat lower for
under-coordinated atoms~\cite{Kotakoski2011b}). As our data shows,
this effect is large enough to overcome the effect of the lowered
target density at the lowest ion energies. While ion energy increases,
a larger fraction of the recoil atoms will receive -- during the
collision -- energies clearly higher than the threshold. Therefore,
the weaker binding is not significant anymore, and the data agrees
better with the simple geometric model. However, at even higher
energies the collision cross section is very small and the time during
which the ion interacts with the target atoms is very short. This
means that the target atom will remain essentially immobile during the
interaction, which correspondingly becomes symmetric over the graphene
plane. Therefore, the transferred momentum will almost completely be
in the in-plane direction. In this case the role of the target density
is negligible after the initial impact, since the displaced target
atom will travel in the in-plane direction as long as it takes to
collide with another target atom. While the overall binding of the
defective membrane is lowered, these secondary collisions can displace
more atoms from the structure than what would happen in the case of a
pristine target. This sets $\gamma \in [0,1]$ also for the highest
irradiation energies, as can be seen in Fig.~\ref{px::slopes}c. Thus,
for energies $K \geq 1$~keV, the modified geometric model can be used
to set the upper and lower bounds for irradiation-induced damage in
graphene.

We stress that defective graphene layers with relatively large defects
are still hardly permeable~\cite{Leenaerts2008} for small atoms and
molecules. Moreover, vacancies in graphene tend to partially ``heal''
themselves by forming non-hexagonal rings due to bond
rotations~\cite{Lee2005b,Kotakoski2011a,Kotakoski2011,Meyer2012}. As
ion beam can give rise to dissociation of atmospheric molecules,
passivation of dangling bonds by hydrogen atoms or incorporation of
foreign atoms (e.g., nitrogen~\cite{Meyer2011}) as substitutional
impurities should also decrease the permeability of graphene
membranes.

In conclusion, we have shown that the response of graphene to ion
irradiation remains consistent to remarkably high vacancy
concentrations (up to 35\%). Although the damaging process varies with
the mass of the impinging ion as well as its energy, we never observed
sudden breakage of the membrane indicating severe structural
instabilities with respect to the ion irradiation. Naturally, whether
the mechanical stability of a perforated graphene membrane is high
enough for a particular application depends on the actual experimental
conditions. However, taking this into account, our results can be used
to predict the effects of ion irradiation on the membrane during the
experiment using the presented simple geometric model in order to
estimate the usability of graphene windows. Our results can also be
used to design optimal conditions for carving nanopores into graphene
using a focused ion beam.

The authors acknowledge financial support by the University of
Helsinki Funds and the Academy of Finland as well as generous grants
of computer time provided by CSC Finland.


\begin{thebibliography}{10}

\bibitem{Novoselov2004}
K.~S. Novoselov, A.~K. Geim, S.~V. Morozov, D. Jiang, Y. Zhang, S.~V. Dubonos,
  I.~V. Grigorieva, and A.~A. Firsov, Science {\bf 306},  666  (2004).

\bibitem{Geim2009a}
A.~K. Geim, Science {\bf 324},  1530  (2009).

\bibitem{CastroNeto2009}
A. {Castro Neto}, F. Guinea, N. Peres, K. Novoselov, and A. Geim, Rev. Mod.
  Phys. {\bf 81},  109  (2009).

\bibitem{Lee2008}
C. Lee, X. Wei, J.~W. Kysar, and J. Hone, Science (New York, N.Y.) {\bf 321},
  385  (2008).

\bibitem{Stolyarova2009}
E. Stolyarova, D. Stolyarov, K. Bolotin, S. Ryu, L. Liu, K. T. Rim, M. Klima,
M. Hybertsen, I. Pogorelsky, I. Pavlishin, {\it et~al.}, Nano Lett. {\bf 9},  332  (2009).

\bibitem{Bunch2008}
J.~S. Bunch, S.~S. Verbridge, J.~S. Alden, A.~M. van~der Zande, J.~M. Parpia,
  H.~G. Craighead, and P.~L. McEuen, Nano Lett. {\bf 8},  2458  (2008).

\bibitem{Jiang2009c}
D.-E. Jiang, V.~R. Cooper, and S. Dai, Nano Lett. {\bf 9},  4019  (2009).

\bibitem{Zhu2011a}
Y. Zhu, S. Murali, M. D. Stoller, K. J. Ganesh, W. Cai, P. J. Ferreira, A.
Pirkle, R. M. Wallace, K. A. Cychosz, M. Thommes {\it et~al.}, Science {\bf 332},  1537  (2011).

\bibitem{Lehtinen2010b}
O. Lehtinen, J. Kotakoski, A.~V. Krasheninnikov, A. Tolvanen, K. Nordlund, and
  J. Keinonen, Phys. Rev. B {\bf 81},  153401  (2010).

\bibitem{Dran2004}
J.-C. Dran, J. Salomon, T. Calligaro, and P. Walter, Nucl. Instr. Meth. Phys.
  Res. B {\bf 219-220},  7  (2004).

\bibitem{Kolmakov2011}
A. Kolmakov, D.~A. Dikin, L.~J. Cote, J. Huang, M.~K. Abyaneh, M. Amati, L.
  Gregoratti, S. G\"{u}nther, and M. Kiskinova, Nature Nanotech. {\bf 6},  651
  (2011).

\bibitem{Tapaszto2008}
L. Tapaszt\'{o}, G. Dobrik, P. Nemes-Incze, G. Vertesy, P. Lambin, and L.
  Bir\'{o}, Phys. Rev. B {\bf 78},  233407  (2008).

\bibitem{Chen2009}
J.-H. Chen, W.~G. Cullen, C. Jang, M.~S. Fuhrer, and E.~D. Williams, Phys. Rev.
  Lett. {\bf 102},  236805  (2009).

\bibitem{Compagnini2009}
G. Compagnini, F. Giannazzo, S. Sonde, V. Raineri, and E. Rimini, Carbon {\bf
  47},  3201  (2009).

\bibitem{Giannazzo2009}
F. Giannazzo, S. Sonde, V. Raineri, and E. Rimini, Appl. Phys. Lett. {\bf 95},
  263109  (2009).

\bibitem{Zhou2010a}
Y.-B. Zhou, Z.-M. Liao, Y.-F. Wang, G.~S. Duesberg, J. Xu, Q. Fu, X.-S. Wu, and
  D.-P. Yu, J. Chem. Phys. {\bf 133},  234703  (2010).

\bibitem{Mathew2011}
S. Mathew, T. Chan, D. Zhan, K. Gopinadhan, A.-R. Barman, M. Breese, S. Dhar,
  Z. Shen, T. Venkatesan, and J.~T. Thong, Carbon {\bf 49},  1720  (2010).

\bibitem{Ney2011}
A. Ney, P. Papakonstantinou, A. Kumar, N.-G. Shang, and N. Peng, Appl. Phys.
  Lett. {\bf 99},  102504  (2011).

\bibitem{Mathew2011a}
S. Mathew, T.~K. Chan, D. Zhan, K. Gopinadhan, A. {Roy Barman}, M.~B.~H.
  Breese, S. Dhar, Z.~X. Shen, T. Venkatesan, and J.~T.~L. Thong, J. Appl.
  Phys. {\bf 110},  084309  (2011).

\bibitem{Compagnini2011}
G. Compagnini, G. Forte, F. Giannazzo, V. Raineri, A.~L. Magna, and I.
  Deretzis, J. Mol. Struct. {\bf 993},  506  (2011).

\bibitem{Chen2011}
J.-H. Chen, L. Li, W.~G. Cullen, E.~D. Williams, and M.~S. Fuhrer, Nature Phys.
  {\bf 7},  535  (2011).

\bibitem{Diaz-Pinto2011}
C. Diaz-Pinto, X. Wang, S. Lee, V. Hadjiev, D. De, W.-K. Chu, and H. Peng,
  Phys. Rev. B {\bf 83},  235410  (2011).

\bibitem{Buchowicz2011}
G. Buchowicz, P.~R. Stone, J.~T. Robinson, C.~D. Cress, J.~W. Beeman, and O.~D.
  Dubon, Appl. Phys. Lett. {\bf 98},  032102  (2011).

\bibitem{Cancado2011}
L.~G. Cancado, A. Jorio, E.~H.~M. Ferreira, F. Stavale, C.~A. Achete, R.~B.
  Capaz, M.~V.~O. Moutinho, A. Lombardo, T.~S. Kulmala, and A.~C. Ferrari, Nano
  Lett. {\bf 11},  3190  (2011).

\bibitem{Ugeda2011}
M.~M. Ugeda, D. Fern\'{a}ndez-Torre, I. Brihuega, P. Pou, A.~J.
  Mart\'{\i}nez-Galera, R. P\'{e}rez, and J.~M. G\'{o}mez-Rodr\'{\i}guez, Phys.
  Rev. Lett. {\bf 107},  116803  (2011).

\bibitem{Ugeda2012}
M. Ugeda, I. Brihuega, F. Hiebel, P. Mallet, J.-Y. Veuillen, J.
  G\'{o}mez-Rodr\'{\i}guez, and F. Yndur\'{a}in, Phys. Rev. B {\bf 85},  121402
   (2012).

\bibitem{Ahlgren2011}
E. \AA~hlgren, J. Kotakoski, and A. Krasheninnikov, Phys. Rev. B {\bf 83},
  115424  (2011).

\bibitem{Lehtinen2011b}
O. Lehtinen, J. Kotakoski, A.~V. Krasheninnikov, and J. Keinonen,
  Nanotechnology {\bf 22},  175306  (2011).

\bibitem{Nordlund1998a}
K. Nordlund, M. Ghaly, R.~S. Averback, M. Caturla, T. {Diaz de la Rubia}, and
  J. Tarus, Phys. Rev. B {\bf 57},  7556  (1998).

\bibitem{Brenner1990b}
D. Brenner, Phys. Rev. B {\bf 42},  9458  (1990).

\bibitem{Brenner1992}
D. Brenner, Phys. Rev. B {\bf 46},  1948  (1992).

\bibitem{Krasheninnikov2001}
A. Krasheninnikov, K. Nordlund, M. Sirvi\"{o}, E. Salonen, and J. Keinonen,
  Phys. Rev. B {\bf 63},  245405  (2001).

\bibitem{Note1}
We refer to the projectile as {\protect \it ion} throughout the text although
  the charge of the impinging particle is not explicitly taken into account in
  our simulation setup.

\bibitem{Ziegler1985}
J.~F. Ziegler, J.~P. Biersack, and U. Littmark, {\em {The Stopping and Range of
  Ions in Matter}} (Pergamon, New York, 1985), .

\bibitem{Berendsen1984}
H. Berendsen, J. Postma, W. {Van Gunsteren}, A. DiNola, and J. Haak, J. Chem.
  Phys. {\bf 81},  3684  (1984).

\bibitem{Tolvanen2007a}
A. Tolvanen, J. Kotakoski, A.~V. Krasheninnikov, and K. Nordlund, Appl. Phys.
  Lett. {\bf 91},  173109  (2007).

\bibitem{Krasheninnikov2002b}
A. Krasheninnikov, K. Nordlund, and J. Keinonen, Appl. Phys. Lett. {\bf 81},
  1101  (2002).

\bibitem{Lehtinen2011}
O. Lehtinen, E. Dumur, J. Kotakoski, A. Krasheninnikov, K. Nordlund, and J.
  Keinonen, Nucl. Instr. Meth. Phys. Res. B {\bf 269},  1327  (2011).

\bibitem{Khare2007}
R. Khare, S.~L. Mielke, J.~T. Paci, S. Zhang, R. Ballarini, G.~C. Schatz, and
  T. Belytschko, Phys. Rev. B {\bf 75},  075412  (2007).

\bibitem{Kotakoski2010}
J. Kotakoski, C. Jin, O. Lehtinen, K. Suenaga, and A. Krasheninnikov, Phys.
  Rev. B {\bf 82},  113404  (2010).

\bibitem{Meyer2012}
J.~C. Meyer, F. Eder, S. Kurasch, V. Skakalova, J. Kotakoski, H.~J. Park, A.
  Chuvilin, G. Benner, A.~V. Krasheninnikov, and U. Kaiser, Phys. Rev. Lett.
	{\bf 108}, 196102 (2012).

\bibitem{Kotakoski2011b}
J. Kotakoski, D. Santos-Cottin, and A.~V. Krasheninnikov, ACS Nano {\bf 6},
  671  (2012).

\bibitem{Leenaerts2008}
O. Leenaerts, B. Partoens, and F.~M. Peeters, Appl. Phys. Lett. {\bf 93},
  193107  (2008).

\bibitem{Lee2005b}
G.-D. Lee, C. Wang, E. Yoon, N.-M. Hwang, D.-Y. Kim, and K. Ho, Phys. Rev.
  Lett. {\bf 95},  205501  (2005).

\bibitem{Kotakoski2011a}
J. Kotakoski, J. Meyer, S. Kurasch, D. Santos-Cottin, U. Kaiser, and A.
  Krasheninnikov, Phys. Rev. B {\bf 83},  245420  (2011).

\bibitem{Kotakoski2011}
J. Kotakoski, A.~V. Krasheninnikov, U. Kaiser, and J.~C. Meyer, Phys. Rev.
  Lett. {\bf 106},  105505  (2011).

\bibitem{Meyer2011}
J.~C. Meyer, S. Kurasch, H. J. Park, V. Skakalova, D. K\"unzel, A. Gro\ss, A.
Chuvilin, G. Algara-Siller, S. Roth, T. Iwasaki {\it et~al.}, Nature Mater. {\bf 10},  209  (2011).

\end{thebibliography}

\end{document}